\documentclass[12pt,showpacs,preprintnumbers,amsmath,amssymb,superscriptaddress,floatfix]{iopart}

\usepackage{graphicx}
\usepackage{dcolumn}
\usepackage{bm}
\usepackage{longtable}

\usepackage{hyperref}

\usepackage[usenames]{color}

\begin{document}

\title{First joint Gravitational Waves search by the AURIGA-EXPLORER-NAUTILUS-Virgo collaboration}
\bigskip

\author{
F.~Acernese$^{12}$, 
P.~Amico$^{20}$, 
M. Alshourbagy$^{21}$, 
F. Antonucci $^{24}$, 
S.~Aoudia$^{13}$, 
P. Astone$^{24}$, 
S.~Avino$^{12}$,  
D.~Babusci$^5$,
L. Baggio$^1$, 
G.~Ballardin$^2$
% ...
}

\address{$^1$Laboratoire d'Annecy-le-Vieux  de physique des particules (LAPP), IN2P3/CNRS, Universit\'e de Savoie, BP 110, F-74941, Annecy-le-Vieux, CEDEX, France;}
\address{$^2$European Gravitational Observatory (EGO), Via E. Amaldi, I-56021 Cascina (PI) Italy;}
% ...
\address{$^5$ INFN, Laboratori Nazionali di Frascati, Frascati, Italy}
% ...
\address{$^{12}$INFN - Sezione di Napoli and/or Universit\`a di Napoli "Federico II" Complesso Universitario di Monte S. Angelo Via Cintia, I-80126 Napoli, Italy and/or Universit\`a di Salerno Via Ponte Don Melillo, I-84084 Fisciano (Salerno), Italy;}
\address{$^{13}$Department Artemis - Observatoire de la C\^ote d'Azur, BP 42209, 06304 Nice Cedex 4, France;}
% ...
\address{$^{20}$INFN Sezione di Perugia and/or  Universit\`a di Perugia, Via A. Pascoli,  I-06123 Perugia - Italy;}
\address{$^{21}$INFN - Sezione di Pisa and/or Universit\`a di Pisa, Via Filippo Buonarroti, 2 I-56127 PISA - Italy;}
% ...
\address{$^{24}$INFN, Sezione di Roma  and/or Universit\`a "La Sapienza",  P.le A. Moro 2, I-00185, Roma, Italy;}
% ...

\bigskip 
\address{E-mail: Lucio.Baggio@lnl.infn.it, guidi@uniurb.it, virginia.re@lnl.infn.it}

\date{\today}

\begin{abstract}
We present results of the search for coincident burst excitations over a 24
hours long data set collected by AURIGA, EXPLORER, NAUTILUS and Virgo
detectors during September 2005. The search of candidate triggers was
performed independently on each of the data sets from single detectors. We
looked for two-fold time coincidences between these candidates using an algorithm
optimized for a given population of sources and we calculated the efficiency
of detection through injections of templated signal waveforms into the streams
of data. To this purpose we have considered the case of signals shaped as
damped sinusoids coming from the galactic center direction. In this framework
our method targets an optimal balance between high efficiency and low false alarm rate, aiming at setting confidence intervals as stringent as possible in terms of the rate of the selected source models.

\end{abstract}

\maketitle

\section{Introduction}

\begin{figure}
\includegraphics[width=150mm]{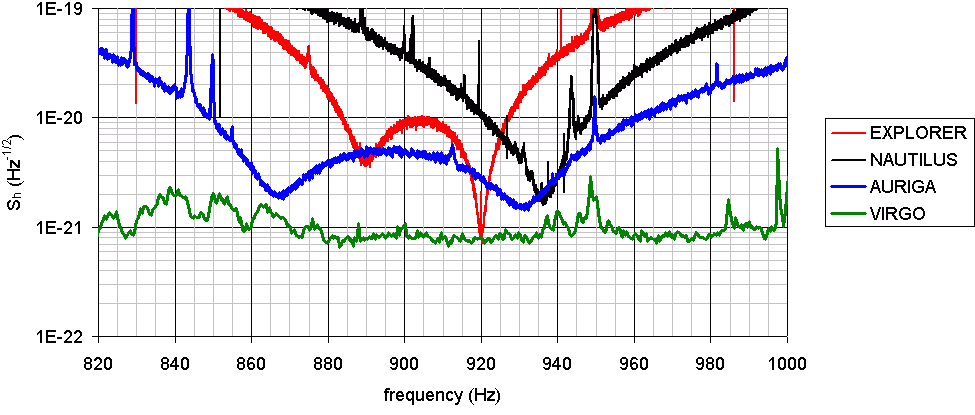}
\centering
\caption{\label{fig:Shh} Typical spectral density of calibrated noise for the three resonant bar detectors during 2005 and for the Virgo interferometer in Sept 2005.}
\end{figure}

The  network composed by the three gravitational wave (GW) resonant detectors AURIGA\cite{3-modes,AURIGA}, EXPLORER and NAUTILUS \cite{ROGa,ROGb} and the interferometer Virgo\cite{Virgo} (hereafter called Virgo--bars network) is heterogeneous, as its single components differ for spectral sensitivity (see figure~\ref{fig:Shh}) and antenna pattern.

In the past years, various searches for GW signals have been independently
performed by individual detectors or by networks of resonant bars (IGEC,
~\cite{IGEC-PRL, IGEC-PRD}, ROG ~\cite{ROG}) or interferometers~\cite{LIGO2004,LIGO2005,LIGO2006}.
In the latter cases the networks were homogeneous: almost same antenna pattern (neglecting a small misalignment), similar (within a factor 2) integrated sensitivity and roughly same observed frequency range (or detection bands).
Therefore, a GW burst would produce approximately the same response in all the detectors of the network (notably, irrespective of direction and polarization of the source). In such cases the magnitude of observed signals can be compared directly.

Previous burst searches among detectors
with  different spectral sensitivity and orientation were 
performed by the TAMA and LIGO Scientific  
Collaboration~\cite{LIGO-TAMA-S2} among interferometers and by the AURIGA and LIGO Scientific Collaborations~\cite{AURLIGOgwdaw,AURLIGOamaldi6} between interferometers and a resonant bar.

The proposed network search strategy for the Virgo--bars data analysis takes
as a starting point the IGEC coincidence search for burst GW events. This
search was innovative with respect to previous searches as it preserved the
detection efficiency by selecting the detectors which, time to time, had
comparable directional sensitivity for sources located at a given sky
position. 
In that case, however, there was no optimization on detection efficiency and
the analysis relied on identical antenna patterns for the detectors. Instead,
for the Virgo--bars network, it is necessary to further develop the idea
included in the IGEC strategy. The detection efficiency will be determined by
studying the software injections (Mock Data Challenge, in the following
referred to as MDC) of a given collection of target waveforms. The approach
attempted in this work is to use the efficiency computation both to tune the
analysis parameters and to calibrate the final results, a step missing in
ref.~\cite{IGEC-PRD}.

The results we present here are obtained in the simpler case of fixed time
coincidence windows and two-fold coincidences among different detectors pairs.
 The coincident counts, divided by the detection efficiency and by the observation time, become then observed rates (or upper limits on rates) for that particular source population.
The relevance of this study is methodological due to the short observation time, the uncertainty on the detectors calibrations and to some approximations in the production of the MDC.

 The paper is organized as follows: in section 2 we introduce the target GW
 signals  and the source population we are dealing with. An overview of the
 exchanged data is presented in section 3. Section 4 presents the results
 obtained from software injections of GW signals into the data and the related
 estimates of the single detectors detection efficiencies and time errors. The coincident search strategy adopted in this work as well as the background estimation method and confidence of detection are described in section 5. The results and final remarks are presented in section 6 and 7 respectively. Finally, we report in Appendix A a summary  of the pipeline main steps and in Appendix B a complete calculation of the energy budget associated to the injected signals. 

\section{Target signals}
The class of transient GW signals 
is extremely large; moreover, such signals may be generated by a large variety of astrophysical sources. 
In this scenario, we have chosen to constrain the source population to the ensamble  
 of waveforms that can be analytically described as a damped sinusoid (DS) with central frequency ranging within the bars
 bandwidths (850-950 Hz) and characterized by decaying times spanning at most a few tens of milliseconds.
This choice is due to the different spectral densities of the various detectors in the collaboration (see figure~\ref{fig:Shh}), so that the interesting signals for our specific network are the ones whose power is concentrated in the bars most sensitive frequency range.

A typical damped sinusoid (DS) waveform is described by the following template:

\begin{equation}
\label{eq:1}
u(t) \propto e^{ - t/\tau } \cos (2\pi f_0 t + \varphi _0 )
 \end{equation}

where $f_0$ is the central frequency and $\tau$ the damping time. These signals can be produced for instance by a ring-down phase following the merger of two black holes \cite{BH}. 
Other sources whose emission can be modelled by (\ref{eq:1}) are f-modes from
neutron stars. The f-modes could produce a wave with variable frequency and
damping time, which may sweep inside the observed frequency band \cite{f-modes}.
 To make a realistic detection, the energy release should be about $10^{-3}-10^{-4}~M_\odot$ for a galactic event (see \ref{norm}).

The astrophysical model for our source population considers elliptically polarized signals (as sources angular momenta should have random directions with respect to the line of sight to the earth) incoming from the Galactic Center.

\section{Overview of exchanged data}
\label{sec:data}

The exchanged data consists of event lists corresponding to 24 hours of data taking, starting from GPS time 810774700, or UTC time 14 Sep 2005 23:11:27. This choice corresponds to the longest scratches of continuous acquisition during the so called ``C7'' run of Virgo, when AURIGA, EXPLORER and NAUTILUS where in stable operation.

Each group exchanged the triggers found  above a chosen threshold by their respective burst event search algorithms. No further tuning of parameters and amplitudes has been done at this stage: a cut based on the magnitude of the events can be optimally set up afterwards based on the relative performances of the detectors at any given time. This selection reduces the number of background events without severely affecting the efficiency for a specific injection class.

Before exchanging all the data, the time information has been offset by a
secret time shift within each group. This was done in order to prevent any
bias which might arise by looking at the zero-delay coincidence counts in the
tuning phase of the analysis.

It has to be noticed that the amplitudes may suffer from a systematic error due to the calibration uncertainty of each detector. This error is declared to be at most $\sim 30\%$ for  Virgo, $\sim 20\%$ for EXPLORER and NAUTILUS, and $\sim 10\%$ for AURIGA.

\subsection{Event Trigger Generators}
From the Virgo side, Power Filter\cite{PF} was the chosen Event Trigger Generator.
Power Filter searches on whitened data for a power excess using different time analysis windows and different frequency
bands and it uses as an indicator of the signal magnitude the (logarithmic) Signal to Noise Ratio (SNR).
Events were exchanged at (logarithmic) $SNR\geq3.4$.

AURIGA group has successfully tested on its data an implementation of WaveBurst Event Trigger Generator, which is an excess power algorithm based on the wavelet transform developed by the LIGO Scientific Collaboration \cite{Waveburst}; 
the exchange threshold was set at amplitude $SNR\geq4.5$.
For the NAUTILUS and EXPLORER detectors the Event Trigger Generator is an adaptive linear filter matched to the impulse response \cite{ROG1,ROG2}. The amplitude calibrated for the impulse response and the SNR were exchanged for each event and the exchange threshold was fixed at $SNR\geq3.8$.

\begin{figure*}
\vspace*{-10mm}
\includegraphics[width=17.5cm]{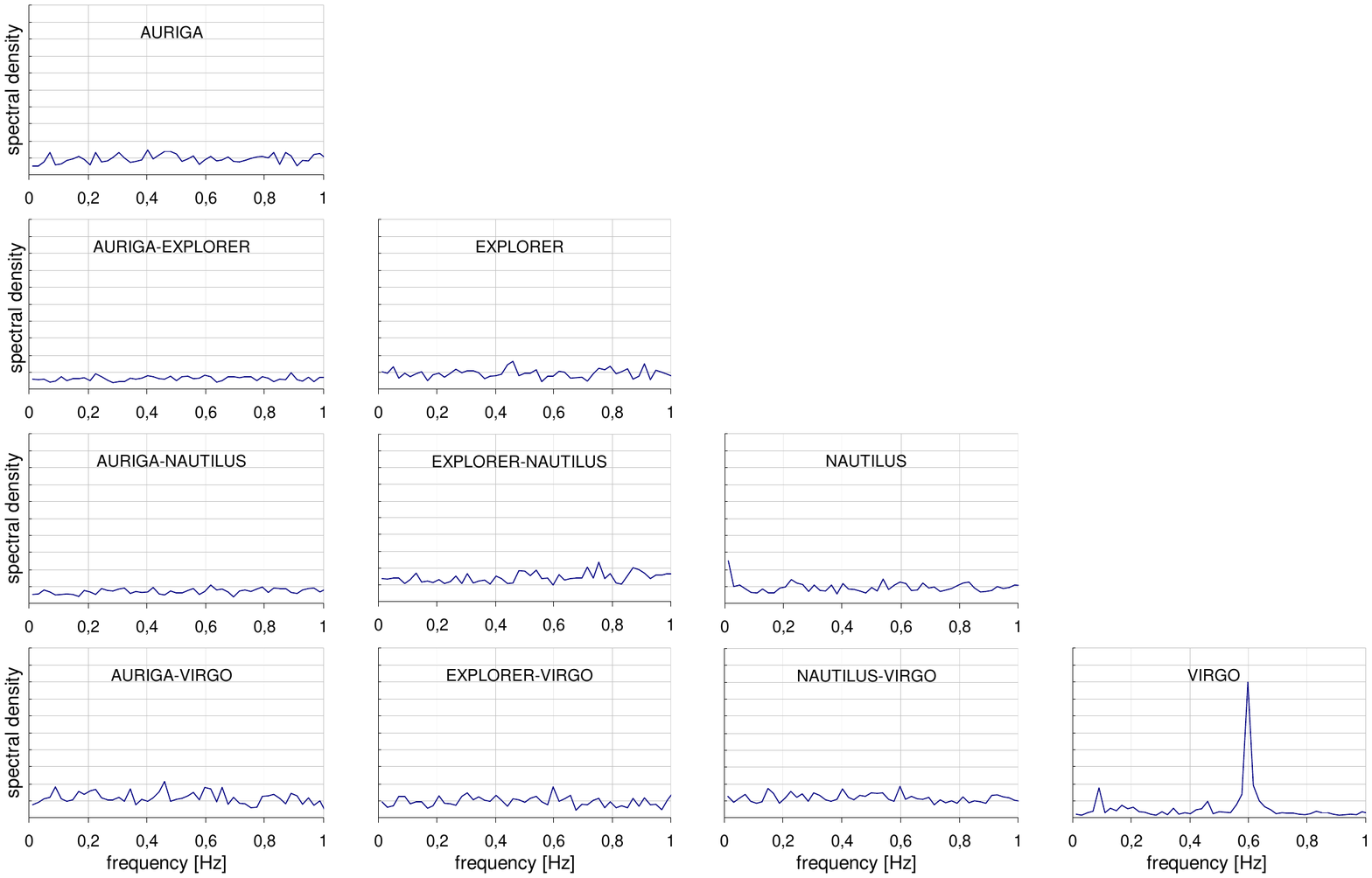}
\vspace*{-25mm}
\caption{\label{fig:correlograms-FT} Spectral densities of correlograms and cross-correlograms for AURIGA, EXPLORER, NAUTILUS and Virgo. The spectra are flat within errors for all cases but for Virgo, where a 0.6Hz noise line is dominant (this is visible as periodic ripples in the Virgo auto-correlogram}
\end{figure*}

\subsection{Data quality: the correlograms}

The histogram of time difference among the outputs of the local Event Trigger Generators (or \textit{correlogram}) with a given bin size is equivalent to counting coincidences with fixed window as a function of the time delay. The absence of cross-correlation is a useful hint that the coincidences at some delay may be considered as representative of the accidental coincidence probability at zero delay.

While single detectors, like Virgo, show some level of auto-correlation in the correlogram of the background events, on the other hand the cross-correlogram is flat as expected for random Poisson point processes.

We found no evidence of modulation in the cross-correlation histogram up to
$\pm$400~s. Looking at the histograms in the Fourier domain (see figure \ref{fig:correlograms-FT}), the spectral density is flat for all detectors and detector couples, with the exception of Virgo, which presents a known strong peak at 0.6~Hz; this is the fundamental pendulum mode of the Virgo mirror suspension, whose excitation is only partially suppressed by the interferometer control loops.

\section{Results from software injections}

The injected GW signals consist of time series of sampled DS with $f_0$ within the sensitive frequency region for
the resonant detectors, $\tau$ spanning between 1 and 30~ms (see table~\ref{tab:MDCs}) and random elliptical polarization.

\begin{table}[b]
\centering
\caption{\label{tab:MDCs} The parameter space of DS signals is described by their central frequency $f_0$ and their decay time $\tau$. The latter takes values spaced logarithmically by about a factor 3, while the frequency axis is sampled at a special subset of frequencies, which was chosen on the basis of the typical narrow-band power spectral densities of the bars.}
\begin{tabular}{c|c}
 $\tau$ (ms)&  $f_0$ (Hz)\\\hline
 1      &             914\\
 3      &	882                          946\\
 10     & 866                898             930\\
 30     & 866  874     906                   930  938\\
\end{tabular}
\end{table}

The source location is chosen at the Galactic Center. For
each detector, a specialized time series is produced including the time delays
and the amplitude attenuation due to antenna pattern (see for example the
amplitude modulation on figure~\ref{fig:optimization}), using the SIESTA
simulation software \cite{siesta:98}.
The simulated signals arrive at the Earth center
approximately evenly spaced by 10~s (with a random jitter of
$\pm$0.5~s), producing a set of 8640 injections over the 24 hrs observation time.

In the following, when referring to the ``amplitude'' of the population, we mean the absolute $h_{rss}$ amplitude of the wave, i.e. the amplitude at the earth of the unprojected wave tensor (see \ref{norm}).
The generated waveform amplitudes for the coincidence analysis range between
$h_{rss} = 10^{-20}$ Hz$^{-1/2}$ and $10^{-18}$ Hz$^{-1/2}$ in order to span the
curve of efficiency vs $h_{rss}$ for all the detectors -- see section
  4.3. 

We acknowledge a coarse approximation in the algorithm that calculates the samples amplitudes for the MDC, resulting in an overall underestimation of the injected signal strength $h_{rss}$ with respect to the declared value. 
The effect of such approximation is fully negligible for the most energetic
signals, while it may cause a spurious loss of efficiency at the lowest
amplitudes and for the less sensitive detectors. Nevertheless, the reported
results are conservative and the methodological relevance of this work is not affected.

\subsection{Time errors}
\label{sec:TimeErrors}
The timing error of all search algorithms is heavily dominated by systematic biases. This is typical of algorithms that are not matched to the particular signal one is looking for. 
For instance, for the Virgo detector, the Power Filter filter bandwidth is $\sim 100$~Hz in the narrowest channel, and the time of arrival is determined by the time when the signal reaches its maximum amplitude. Because of this the biases for $1$~ms and $10$~ms long DS are $0.8$~ms and $3.6$~ms respectively. 

For the AURIGA detector, the bandwidth is narrower, causing a larger distortion of the signal. Moreover the time associated to the event is computed as the baricenter of the signal profile above threshold. The amount of the bias depends strongly on the time duration and also on the central frequency of the signal. Altogether, for the durations $ \leq 30$~ms the bias for DS ranges from $\sim 3$~ms to $\sim 30$~ms depending on the central frequency \cite{Drago-efficiency}.

The linear filter matched to a delta signal is unbiased for wide band signals,
such as $1$~ms long DS, but it is more biased as the signal duration gets longer \cite{ROG3}. For the considered DS the worst time error of EXPLORER and NAUTILUS was of the order of $8$~ms and $16$~ms for damping time $\tau=10$~ms and $\tau=30$~ms respectively.

\begin{figure*}
\centering
\includegraphics[width=5cm]{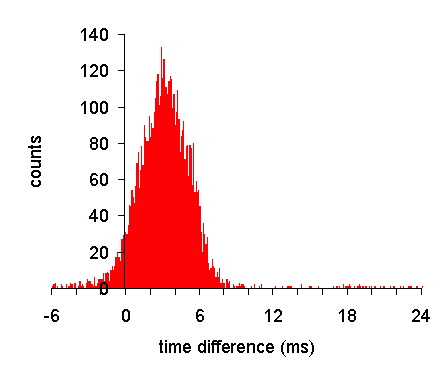}
\includegraphics[width=5cm]{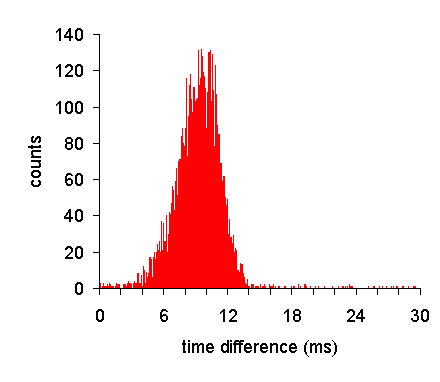}
\includegraphics[width=5cm]{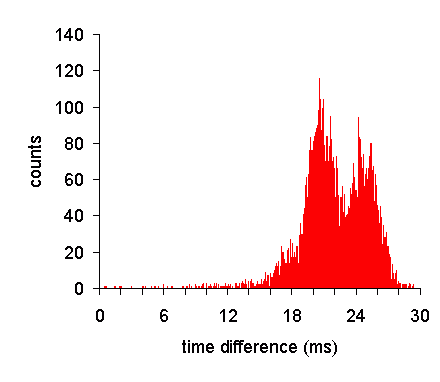}
\caption{\label{fig:times} Distribution of time differences of detected events in coincident injection of DS(914Hz;1ms) (\textit{left}),  DS(930Hz;10ms) (\textit{center}) and DS(930Hz;30ms) (\textit{right}), at $h_{rss}=10^{-18}$ Hz$^{-1/2}$ for the couple AURIGA-Virgo.}
\end{figure*}

In figure~\ref{fig:times} we see for example that the maximum time difference between AURIGA and Virgo is $\leq$ 30 ms.

\subsection{Distribution of amplitudes of accidental events and injected signals}

\begin{figure*}
\centering
\includegraphics[width=75mm]{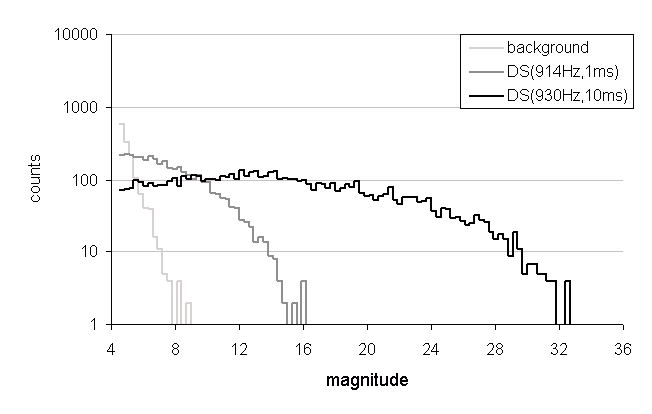}
\includegraphics[width=75mm]{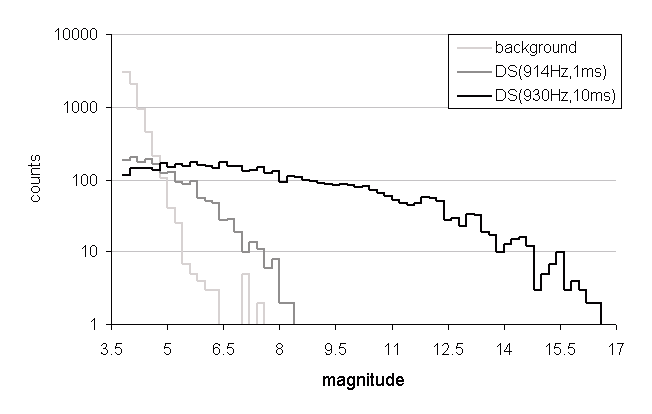}\\
\includegraphics[width=75mm]{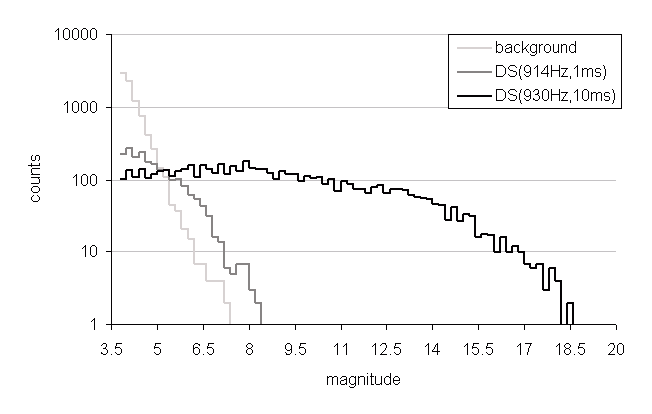}
\includegraphics[width=75mm]{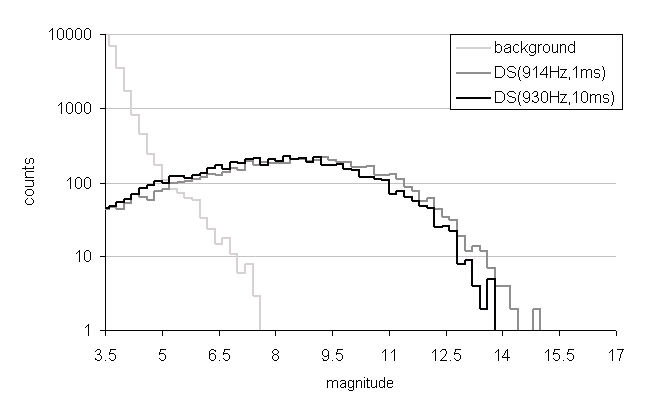}
\caption{\label{fig:data} Distribution of detected ``event magnitudes'' for
  background events and injections of DS(914Hz;1ms) and DS(930Hz;10ms) at
  $h_{rss}=10^{-19}Hz^{-1/2}$ for AURIGA (\textit{top-left}), EXPLORER
  (\textit{top-right}), NAUTILUS (\textit{bottom-left}) and Virgo
  (\textit{bottom-right}). The "magnitude" can be the SNR ratio given by
  WaveBurst algorithm (AU), the one given by a linear matched filter (EX and
  NA) and the logarithmic SNR of Power Filter (Virgo). Given the very different meaning of these quantities, the plots obtained for different detectors cannot be compared directly.}
\end{figure*}

The single detector search algorithms provide different estimates of the magnitude of the signal. Although not directly comparable among different detectors, the event magnitudes provided by each detector show how much the population of injected waveforms stems from the noise distribution. An example can be seen in figure~\ref{fig:data}.

\subsection{Efficiency of detection}

\begin{figure*}
\centering
\includegraphics[width=8.4cm]{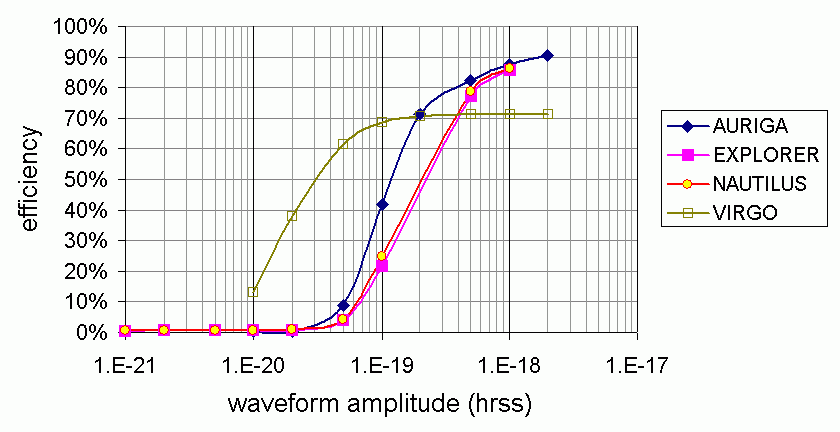}
\includegraphics[width=6.7cm]{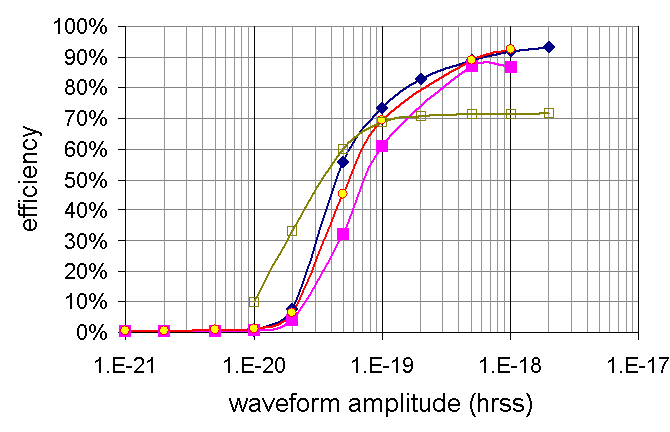}
\caption{Efficiency of detection for the four detectors when recovering injections of DS(914Hz;1ms) (\textit{left}) and DS(930Hz;10ms) (\textit{right}) at different values of $h_{rss}$. In the selected 24 hours, $\sim 7$ hours of Virgo data have been excluded by epoch vetoes based on data quality. That is why efficiency levels at about 70\%.}
\label{fig:effic} 
\end{figure*}

The software injections have been used also to monitor the detection efficiency of the single detectors. 
The efficiency is computed for different waveform amplitudes and using a $\pm 40$~ms time window around the injection times: we look for coincidences between the injected signals and the events found by the Event Trigger Generators.
The calculation of the efficiency is based on the nominal 24 hours allocated for this search. Consequently, dead times in the data due, for example, to epoch vetoes affect the average efficiency.
This is the case for detector Virgo: its duty time is, in fact, far less than
100\% in the 24 hours considered (about 7 hours are vetoed out because of bad
data quality), which finally reduces the attainable average efficiency (see
figure \ref{fig:effic}).
The resonant detectors have instead a very stable duty cycle and they show improved performances with respect to Virgo when the signal is fully contained in their bandwidth (which requires special selection of central frequency and long signal duration).

In conclusion, our definition of efficiency includes also dead times when data are missing or vetoed out from the nominal set, as we are interested in whether or not the network was able to recover the injected signals. We will comply with this comprehensive definition of efficiency throughout the paper.

\section{Coincidence analysis}

A coincidence between two detectors is defined as the fulfilling of the relation

\begin{equation}
\left| {t_{j_k }^{(k)}  - t_{j_h }^{(h)}  - \Delta t_{hk} } \right| < T_w
\end{equation}

where $t_{j_k }^{(k)}$ is the estimated time of arrival of the $j_k$ signal in the detector labelled $k$ and $\Delta t_{hk}$ is the time of flight between the sites of the two detectors.  For the case of our network and for signals coming from the Galactic Center, the maximum time delay ranges between 0.3 ms for the couple Virgo-AURIGA and 2.4 ms for EXPLORER-NAUTILUS.
From the results of the MDC injections (specifically,
sec.\ref{sec:TimeErrors}), we set in the following $T_w = 40$~ms. After the tuning of the analysis (which will be described in the following sections), we checked, for all the configurations shown on table\ref{tab:results}, that the loss in the overall efficiency with this coincidence time window is at most $1\%$.

\subsection{Thresholds optimization}

As we can see from figure~\ref{fig:data}, for a small increase in the
magnitude threshold a large reduction of the background counts can be
expected. The accidental coincidence rate between two detectors is
proportional to the event rate of the two detectors, so we can act on one of
the two thresholds, or on both. 
The trade off is the reduction of detection efficiency. Sometimes the detected
magnitude of the injected events is large enough to increase the threshold up
to exclude any background coincidence, while in other time periods lower
thresholds are preferred to preserve the detection efficiency.

In order to quantify these statements, we consider a gain function defined as the 
ratio between the average efficiency  and the square root of the background counts. 
The rationale for this choice derives from the procedure to set confidence intervals on the number of true coincidences. In fact, the background of accidental coincidences can always be subtracted from the found (total) number of coincidences. The residual of the subtraction is --loosely speaking-- the number of truly correlated events between the detectors. In this sense, our gain function is the ratio between the average efficiency and the fluctuations of the background. 
 As our source population is set in the Galactic Center, we apply a time
 varying threshold (calculated every 30 minutes by maximizing the gain
 function, as above stated). This implies that for each time bin the threshold
 is set at the level corresponding to the maximum of the gain function 
for the pair of detectors.
The overall result is that we apply over the entire data set and for each detector of the couple a non-constant cut on the event magnitude, using a  
threshold set every 30 minutes (the analysis
  pipeline is discussed in more detail in Appendix A).

Figure~\ref{fig:optimization} shows an example of such adaptive thresholds. As Virgo, in this example, can see the injected events much better than AURIGA, the algorithm starts with raising Virgo threshold. This cleans up most of the coincident events. The efficiency of AURIGA is therefore preserved, as its threshold is left almost untouched.

\subsection{Background and efficiency of the network}

The efficiency of detection is empirically defined by the sets of data containing MDC injections. The ratio between the coincident events due to injections found in the detector couple under study and the known number of injected events gives the empirical estimate of the efficiency \footnote{The contamination due to accidental coincidences was found to be negligible.}.

For the background estimation, we first take care of the (possible) true
correlated events present in the data by shifting the times of the event lists
of different detectors before looking for
coincidences in time. By repeating this operation a number of times, we get renewed instances of the 
counting experiment:  we will refer to this procedure as \textit{time
  delay analysis}.
 Altogether, the coincidences from hundreds of shifted
configurations provide a rich population from which we can determine the main
parameters of the background distribution. If the time slide measurements are independent from each other, 
the number of accidental coincidences in each time shift should be 
Poisson distributed. We tested this hypothesis by means of 
a $\chi^2$ test just on those searches which have an high expected number of accidental
coincidences, $N_b$, so
to ensure a sufficiently large data sample. 
The corresponding p-values were  not inconsistent with the 
Poisson model for the expected number of accidentals. 

The optimization procedure described in the previous section determines the cuts on the data set based on a function of the estimated background and efficiency. Hence these two estimates will be biased, sometimes severely, and cannot be used for setting confidence intervals. For this reason, we preliminarily divide the original data (accidental coincidences and MDC injections) into two equal size subsets: one is used in the optimization phase. The threshold levels obtained at the end of the optimisation are subsequently applied to the second halves of the data, without further tuning. Background counts and efficiency are thus computed from this second subset, giving unbiased estimates.

\begin{figure*}
\centering
\includegraphics[width=175mm]{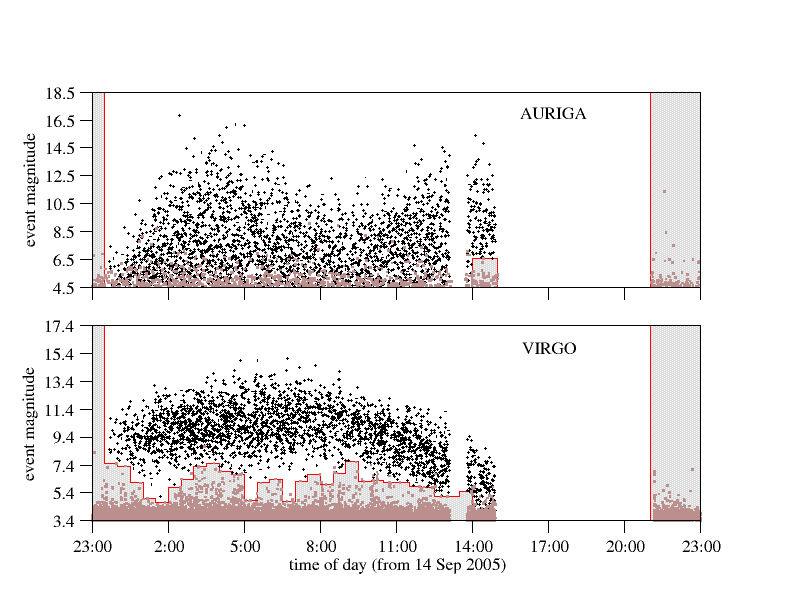}
\caption{\label{fig:optimization} Example of threshold placing for detection
  optimization for the template DS(914Hz;1ms) at $h_{rss}=10^{-19}$ Hz$^{-1/2}$
  for the couple AURIGA (\textit{above}) and Virgo (\textit{below}). The
  abscissa represents the period of the exchanged 24 hours of data taking, the
  ordinate is the event magnitude given respectively by the WaveBurst and the
  PowerFilter algorithms. The light markers indicate background coincident
  events, the darker ones are events generated by injections. The
  modulation of the event magnitude by the antenna pattern is clearly visible
  in the two plots. The events form
  coincident couples, i.e. to each event in AURIGA the corresponding event in
  Virgo is plotted. Excluding one event from one detector automatically
  excludes the paired one in the other. Thus, given the relatively good
  separation of MDC events from background events in Virgo, the threshold here
  is raised up to the limit where basically all background events are
  excluded, while the threshold in AURIGA is left almost at its initial value.}
\end{figure*}

\subsection{Setting confidence intervals}
\label{sec:confidence}

We first set the confidence interval on the number of correlated events detected in coincidence, following a unified approach in the spirit of ~\cite{feldmancousinsPRD}. However, the procedure we adopt to build the confidence belt is different and its fundamentals has been discussed in ~\cite{IGEC-PRD} and ~\cite{PHYSTAT2003}.
We start by considering the likelihood of the number of coincidences at zero-delay $N_c$ as a function of the expected values of the accidental counts and of the correlated events, $N_b$ and $N_{GW}$ respectively:

\begin{equation}
l(N_c ; N_b, N_{GW}) = \frac{(N_b + N_{GW})^{N_c}}{N_c !} e^{-(N_b +N_{GW})}
 \end{equation}

The confidence intervals on $N_{GW}$ are built by integrating the likelihood

 \begin{equation}
I=\frac{\left[ \int_{N_{inf}}^{N_{sup}} l(N_c ;N_b , N_{GW})dN_{GW}\right]}{\int_0^\infty l(N_c ; N_b , N_{GW}) dN_{GW}}
 \end{equation}

to find the smallest interval $[ N_{inf}, N_{sup}] $ corresponding to the chosen $I$ value. The set of these intervals computed for the possible $N_c$ values makes the wanted confidence belt of $N_{GW}$ vs $N_c$ for fixed level of background $N_b$. It is well known that the $I$ cannot be interpreted as a frequentist probability; therefore, the actual probability that the confidence intervals $[N_{inf},N_{sup}]$ include the true $N_{GW}$ value, i.e. the coverage, must be empirically determined by a Monte Carlo~\cite{PHYSTAT2003}. The quoted coverage is the minimum coverage ensured by the belt \footnote{Notice that at $95\%$ coverage, the upper limit set by this belt is  3.6 counts when $N_c = 0$, regardless of the value of the background, $N_b$. }. 

In the present work, we modify the above procedure by adding a more stringent test of the null hypothesis in order to increase the coverage when $N_{GW} = 0$. In practice, we want a more stringent false alarm to issue a two-sided confidence interval with lower extreme greater than zero. This is done by performing a Poisson one-tail test on the found coincidences $N_c$ assuming the null hypothesys, $N_{GW} = 0$. The significance of the test is set at a higher level than the coverage of the belt, as discussed in the next section. In case the test is passed, the lower extreme of the confidence interval $N_{inf}$ is extended to zero. 

\begin{figure}
\centering
\includegraphics[width=12cm]{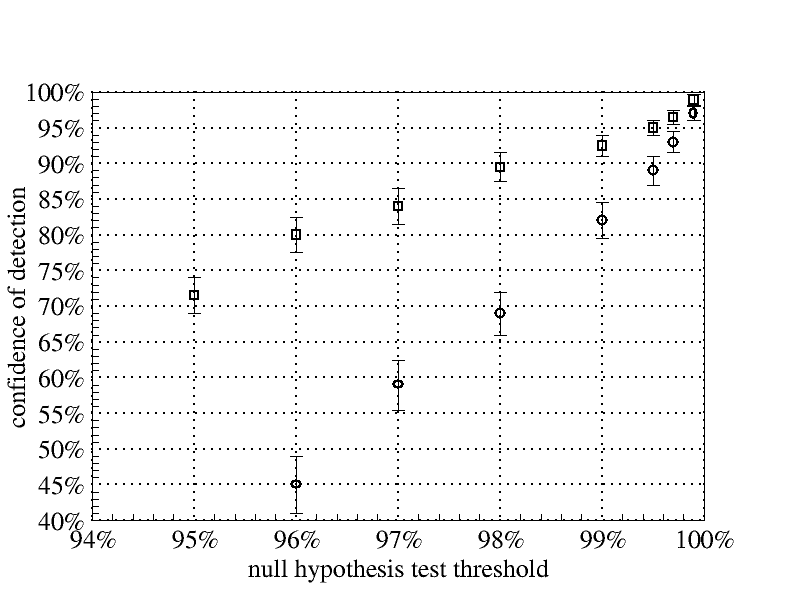}
\caption{\label{fig:confidence} Measured global confidence of detection over all configurations \textit{vs} nominal confidence of the null hypothesis test for each single search. The circles refer to the whole list of trials, the squares to the configurations which gave the best upper limits.}
\end{figure}

\subsection{Multiple trials and global confidence}
\label{mult}
In this analysis we perform many searches for different signal waveforms and
amplitudes by different detector pairs. The trial factor is therefore large
and we are interested in controlling the global probability of false claim of
the whole experiment. To estimate it, we considered the 400 different time-delayed configurations obtained by time shifting the original data set, assuming they are independent realizations of the experiment with no correlated events. The resulting coincidences has to be accidental and we can empirically estimate the distribution of these coincidence counts. We can then determine the global confidence of detection of the entire experiment as a function of the chosen significance of the null hypothesis test on each single trial. The results are shown as circle data points in figure~\ref{fig:confidence}. In particular, to have a global confidence of detection of at least $95\%$, we need to set the significance up to $\sim 0.999$ on the null hypothesis test for each single trial.

There is a trade off between the global false claim probability and the detection efficiency of the experiment: to decrease the false claim probability it becomes more difficult to recognize a true signal. For instance, with $N_b \sim 0.1$, as observed in most of the configurations, and the threshold of the 1-tail Poisson test set at 99.9\%, we need at least 3 coincidences detected in a couple of detectors in order to reject the null hypothesis and make a claim for correlated events.

In order to reduce this drawback, we attempted to limit as much as possible the trial factor of the experiment. In particular, for each signal waveform and amplitude we considered only the results produced by the pair of detectors which was performing best (square data points in figure~\ref{fig:confidence}). To select the best performing pair, we
compute fake upper limits per each pair under the assumption that the number of found coincidences is compatible to the expected background of the pair. Only the pair producing the more stringent fake upper limit is then searched for true coincidences. In this way, setting a significance of $0.999$ on the null hypothesis test for each single trial, we achieve a global confidence of $\sim99\%$, i.e. a false claim probability of $\sim1\%$. 

\begin{figure*}
\centering
\includegraphics[width=12cm]{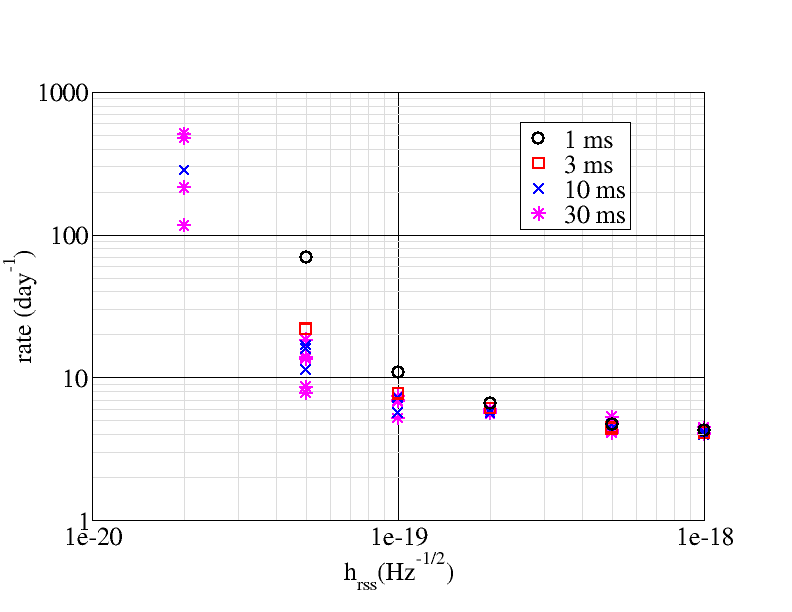}
\caption{\label{fig:results} Upper limit at $95\%$ conservative coverage on the event rate as a function of the population amplitude. Upper limits referring to the same duration of the signal but different central frequencies are grouped with the same symbol. Notice that the systematic error of about 10-30\% on the amplitude calibration has not been taken into account.}
\end{figure*}

\section{Results}
\label{results}
As a last step, we searched for coincidences at the true time in all the selected configurations.
The final outcome was consistent with no rejection of the null hyphothesis test at 99.9\% confidence for each configuration, corresponding to a global false claim probability of 1\%. 
The confidence interval are, hence, upper limits on the rate of incoming GWs.

Detailed results are presented in
  table~\ref{tab:results}, where, for each injected waveform, the estimated efficiency, the average background, the zero-lag coincidence counts and the corresponding $95\%$ upper limit are reported. 
Rescaling the listed efficiencies by their asymptotical values, we can easily infer that, for the 11 injected waveforms, the so-called $h_{rss}^{50\%}$ \footnote{The signal amplitude with  $50\%$ detection 
probability. } ranges from $5\cdot 10^{-20}$ Hz$^{-1/2}$ (for the DS with central frequency $f_0$=930 Hz and $\tau$=30 ms for the AURIGA-NAUTILUS pair) to $1\cdot 10^{-19}$ Hz$^{-1/2}$ (for the DS with central frequency $f_0$=914 Hz and $\tau$=1 ms for the AURIGA-Virgo pair). 
We included also a few configurations with low amplitude signals: in order to preserve the residual efficiencies, the related average backgrounds are quite high, leading to non null zero-delay coincidence counts.      

Figure~\ref{fig:results} shows the upper limits as a
function of signal amplitude. The asymptote for large amplitude signals
is inversely proportional to the observation time and to the asymptotical efficiency and depends on the confidence belt: 
for this 24 hr search, it is  $\simeq4.0$~events/day, as our maximum efficiency is $\simeq 90\%$ and the chosen confidence belt (see sec.\ref{sec:confidence}) with $95\%$ C.L. sets a pretty conservative value (3.6 counts).

\begin{longtable}{c|ccc|cc|cc}
\caption{\label{tab:results}
Results of the two-fold coincidence searches, for the chosen
couples of detectors (see sec.\ref{mult}) and for each set of the waveform parameters and amplitudes (Second, third and fourth column). 
The last four columns represent,
respectively, the efficiency of detection measured by means of MDC injections, the
average accidental coincidence counts ($N_b$), the number of coincidences
found at zero-delay ($N_c$) and the corresponding $95\%$ upper limit. The errors associated with
the efficiency and background estimates show the $1\sigma$ statistic fluctuation
(\textit{apices}).
}\\
 \hline \hline
 detector pair  &       $f_0$ (Hz) &    $\tau$ (ms) & $h_{rss}$ (Hz$^{-1/2}$) & efficiency ($\%$)& $N_b$ & $N_c$ & UL\\
 \hline \hline
\endfirsthead
\multicolumn{7}{l}\textit{Table~\ref{tab:results} continued}\\
 \hline
 detector pair  &       $f_0$ (Hz) &    $\tau$ (ms) & $h_{rss}$ (Hz$^{-1/2}$)
 & efficiency ($\%$) & $N_b$ & $N_c$ & UL\\
 \hline \hline
\endhead
\multicolumn{7}{r}\textit{continue ... }
\endfoot
\hline \hline
\endlastfoot
\hline
 \hline
AUR--VIR   &	914 &	1 &	$5\cdot 10^{-20}$ &	$5.1^{\pm 0.3}_{}$ &	$0.046^{\pm 0.011}$ &	0 &	69.2\\
AUR--VIR   &	914 &	1 &	$1\cdot 10^{-19}$ &	$32.6^{\pm 0.7}_{}$ &	$0.093^{\pm 0.016}$ &	0 &	10.9\\
AUR--VIR   &	914 &	1 &	$2\cdot 10^{-19}$ &	$54.0^{\pm 0.8}_{}$ &	$0.101^{\pm 0.016}$ &	0 &	6.6\\
AUR--NAU   &	914 &	1 &	$5\cdot 10^{-19}$ &	$75.2^{\pm 0.7}_{}$ &	$0.068^{\pm 0.014}$ &	0 &	4.7\\
AUR--NAU   &	914 &	1 &	$1\cdot 10^{-18}$ &	$83.0^{\pm 0.6}_{}$ &	$0.071^{\pm 0.014}$ &	0 &	4.3\\
AUR--VIR   &	914 &	1 &	$2\cdot 10^{-18}$ &	$68.1^{\pm 0.7}_{}$ &	$0.103^{\pm 0.017}$ &	0 &	5.2\\
 \hline
AUR--VIR   &	882 &	3 &	$5\cdot 10^{-20}$ &	$16.3^{\pm 0.6}_{}$ &	$0.086^{\pm 0.015}$ &	0 &	21.9\\
AUR--VIR   &	882 &	3 &	$1\cdot 10^{-19}$ &	$45.3^{\pm 0.8}_{}$ &	$0.091^{\pm 0.016}$ &	0 &	7.8\\
AUR--VIR   &	882 &	3 &	$2\cdot 10^{-19}$ &	$58.5^{\pm 0.8}_{}$ &	$0.098^{\pm 0.016}$ &	0 &	6.1\\
AUR--EXP   &	882 &	3 &	$5\cdot 10^{-19}$ &	$79.8^{\pm 0.6}_{}$ &	$0.058^{\pm 0.012}$ &	0 &	4.5\\
AUR--EXP   &	882 &	3 &	$1\cdot 10^{-18}$ &	$86.6^{\pm 0.5}_{}$ &	$0.056^{\pm 0.012}$ &	0 &	4.1\\
AUR--VIR   &	882 &	3 &	$2\cdot 10^{-18}$ &	$68.6^{\pm 0.7}_{}$ &	$0.103^{\pm 0.017}$ &	0 &	5.2\\
 \hline
AUR--VIR   &	946 &	3 &	$5\cdot 10^{-20}$ &	$16.2^{\pm 0.6}_{}$ &	$0.091^{\pm 0.016}$ &	0 &	22.0\\
AUR--VIR   &	946 &	3 &	$1\cdot 10^{-19}$ &	$46.2^{\pm 0.8}_{}$ &	$0.086^{\pm 0.015}$ &	0 &	7.7\\
AUR--VIR   &	946 &	3 &	$2\cdot 10^{-19}$ &	$58.6^{\pm 0.8}_{}$ &	$0.088^{\pm 0.015}$ &	0 &	6.1\\
AUR--NAU   &	946 &	3 &	$5\cdot 10^{-19}$ &	$81.1^{\pm 0.6}_{}$ &	$0.098^{\pm 0.016}$ &	0 &	4.4\\
EXP--NAU   &	946 &	3 &	$1\cdot 10^{-18}$ &	$85.9^{\pm 0.5}_{}$ &	$0.036^{\pm 0.010}$ &	0 &	4.1\\
AUR--VIR   &	946 &	3 &	$2\cdot 10^{-18}$ &	$67.8^{\pm 0.7}_{}$ &	$0.101^{\pm 0.016}$ &	0 &	5.2\\
 \hline
AUR--VIR   &	866 &	10 &	$5\cdot 10^{-20}$ &	$21.0^{\pm 0.6}_{}$ &	$0.078^{\pm 0.014}$ &	0 &	16.9\\
AUR--VIR   &	866 &	10 &	$1\cdot 10^{-19}$ &	$48.5^{\pm 0.8}_{}$ &	$0.073^{\pm 0.014}$ &	0 &	7.3\\
AUR--VIR   &	866 &	10 &	$2\cdot 10^{-19}$ &	$60.2^{\pm 0.7}_{}$ &	$0.063^{\pm 0.013}$ &	0 &	5.9\\
AUR--EXP   &	866 &	10 &	$5\cdot 10^{-19}$ &	$76.2^{\pm 0.6}_{}$ &	$0.071^{\pm 0.014}$ &	0 &	4.7\\
AUR--EXP   &	866 &	10 &	$1\cdot 10^{-18}$ &	$84.3^{\pm 0.6}_{}$ &	$0.053^{\pm 0.012}$ &	0 &	4.2\\
AUR--VIR   &	866 &	10 &	$2\cdot 10^{-18}$ &	$69.2^{\pm 0.7}_{}$ &	$0.103^{\pm 0.017}$ &	0 &	5.1\\
 \hline
AUR--VIR   &	898 &	10 &	$5\cdot 10^{-20}$ &	$22.3^{\pm 0.6}_{}$ &	$0.083^{\pm 0.015}$ &	0 &	15.9\\
AUR--VIR   &	898 &	10 &	$1\cdot 10^{-19}$ &	$49.4^{\pm 0.8}_{}$ &	$0.088^{\pm 0.015}$ &	0 &	7.2\\
AUR--VIR   &	898 &	10 &	$2\cdot 10^{-19}$ &	$60.4^{\pm 0.7}_{}$ &	$0.088^{\pm 0.015}$ &	0 &	5.9\\
AUR--EXP   &	898 &	10 &	$5\cdot 10^{-19}$ &	$82.8^{\pm 0.6}_{}$ &	$0.066^{\pm 0.013}$ &	0 &	4.3\\
AUR--EXP   &	898 &	10 &	$1\cdot 10^{-18}$ &	$87.7^{\pm 0.5}_{}$ &	$0.068^{\pm 0.014}$ &	0 &	4.1\\
AUR--VIR   &	898 &	10 &	$2\cdot 10^{-18}$ &	$69.2^{\pm 0.7}_{}$ &	$0.098^{\pm 0.016}$ &	0 &	5.1\\
 \hline
AUR--VIR   &	930 &	10 &	$2\cdot 10^{-20}$ &	$2.4^{\pm 0.2}_{}$ &	$4.1^{\pm 0.10}$ &	4 &	283.9\\
AUR--VIR   &	930 &	10 &	$5\cdot 10^{-20}$ &	$31.1^{\pm 0.7}_{}$ &	$0.071^{\pm 0.014}$ &	0 &	11.4\\
AUR--NAU   &	930 &	10 &	$1\cdot 10^{-19}$ &	$61.9^{\pm 0.7}_{}$ &	$0.068^{\pm 0.014}$ &	0 &	5.7\\
AUR--VIR   &	930 &	10 &	$2\cdot 10^{-19}$ &	$62.7^{\pm 0.7}_{}$ &	$0.081^{\pm 0.015}$ &	0 &	5.7\\
EXP--NAU   &	930 &	10 &	$5\cdot 10^{-19}$ &	$83.7^{\pm 0.6}_{}$ &	$0.036^{\pm 0.010}$ &	0 &	4.3\\
AUR--NAU   &	930 &	10 &	$1\cdot 10^{-18}$ &	$89.4^{\pm 0.5}_{}$ &	$0.083^{\pm 0.015}$ &	0 &	4.0\\
AUR--VIR   &	930 &	10 &	$2\cdot 10^{-18}$ &	$69.3^{\pm 0.7}_{}$ &	$0.103^{\pm 0.017}$ &	0 &	5.1\\
 \hline
AUR--VIR   &	866 &	30 &	$2\cdot 10^{-20}$ &	$2.1^{\pm 0.2}_{}$ &	$9.1^{\pm 0.15}$ &	11 &	508.1\\
AUR--VIR   &	866 &	30 &	$5\cdot 10^{-20}$ &	$25.5^{\pm 0.7}_{}$ &	$0.021^{\pm 0.007}$ &	0 &	14.0\\
AUR--VIR   &	866 &	30 &	$1\cdot 10^{-19}$ &	$51.9^{\pm 0.8}_{}$ &	$0.013^{\pm 0.006}$ &	0 &	6.8\\
AUR--VIR   &	866 &	30 &	$2\cdot 10^{-19}$ &	$61.8^{\pm 0.7}_{}$ &	$0.053^{\pm 0.012}$ &	0 &	5.8\\
AUR--VIR   &	866 &	30 &	$5\cdot 10^{-19}$ &	$66.5^{\pm 0.7}_{}$ &	$0.051^{\pm 0.012}$ &	0 &	5.3\\
AUR--EXP   &	866 &	30 &	$1\cdot 10^{-18}$ &	$78.8^{\pm 0.6}_{}$ &	$0.068^{\pm 0.014}$ &	0 &	4.5\\
AUR--VIR   &	866 &	30 &	$2\cdot 10^{-18}$ &	$69.2^{\pm 0.7}_{}$ &	$0.111^{\pm 0.017}$ &	0 &	5.1\\
 \hline
AUR--VIR   &	874 &	30 &	$2\cdot 10^{-20}$ &	$2.4^{\pm 0.2}_{}$ &	$9.5^{\pm 0.15}$ &	12 &	476.2\\
AUR--VIR   &	874 &	30 &	$5\cdot 10^{-20}$ &	$26.4^{\pm 0.7}_{}$ &	$0.046^{\pm 0.011}$ &	0 &	13.5\\
AUR--VIR   &	874 &	30 &	$1\cdot 10^{-19}$ &	$51.5^{\pm 0.8}_{}$ &	$0.033^{\pm 0.009}$ &	0 &	6.9\\
AUR--VIR   &	874 &	30 &	$2\cdot 10^{-19}$ &	$61.6^{\pm 0.7}_{}$ &	$0.043^{\pm 0.011}$ &	0 &	5.8\\
AUR--EXP   &	874 &	30 &	$5\cdot 10^{-19}$ &	$73.1^{\pm 0.7}_{}$ &	$0.076^{\pm 0.014}$ &	0 &	4.9\\
AUR--EXP   &	874 &	30 &	$1\cdot 10^{-18}$ &	$82.3^{\pm 0.6}_{}$ &	$0.063^{\pm 0.013}$ &	0 &	4.3\\
AUR--VIR   &	874 &	30 &	$2\cdot 10^{-18}$ &	$69.8^{\pm 0.7}_{}$ &	$0.111^{\pm 0.017}$ &	0 &	5.1\\
 \hline
AUR--VIR   &	906 &	30 &	$5\cdot 10^{-20}$ &	$19.4^{\pm 0.6}_{}$ &	$0.086^{\pm 0.015}$ &	0 &	18.4\\
AUR--EXP   &	906 &	30 &	$1\cdot 10^{-19}$ &	$48.0^{\pm 0.8}_{}$ &	$0.061^{\pm 0.013}$ &	0 &	7.4\\
AUR--VIR   &	906 &	30 &	$2\cdot 10^{-19}$ &	$59.5^{\pm 0.7}_{}$ &	$0.068^{\pm 0.014}$ &	0 &	6.0\\
AUR--EXP   &	906 &	30 &	$5\cdot 10^{-19}$ &	$82.8^{\pm 0.6}_{}$ &	$0.066^{\pm 0.013}$ &	0 &	4.3\\
AUR--EXP   &	906 &	30 &	$1\cdot 10^{-18}$ &	$86.3^{\pm 0.5}_{}$ &	$0.066^{\pm 0.013}$ &	0 &	4.1\\
AUR--VIR   &	906 &	30 &	$2\cdot 10^{-18}$ &	$69.3^{\pm 0.7}_{}$ &	$0.103^{\pm 0.017}$ &	0 &	5.1\\
 \hline
AUR--VIR   &	930 &	30 &	$2\cdot 10^{-20}$ &	$9.3^{\pm 0.4}_{}$ &	$16.3^{\pm 0.20}$ &	16 &	116.2\\
AUR--NAU   &	930 &	30 &	$5\cdot 10^{-20}$ &	$45.4^{\pm 0.8}_{}$ &	$0.043^{\pm 0.011}$ &	0 &	7.8\\
AUR--NAU   &	930 &	30 &	$1\cdot 10^{-19}$ &	$69.0^{\pm 0.7}_{}$ &	$0.056^{\pm 0.012}$ &	0 &	5.2\\
AUR--VIR   &	930 &	30 &	$2\cdot 10^{-19}$ &	$64.0^{\pm 0.7}_{}$ &	$0.023^{\pm 0.008}$ &	0 &	5.6\\
EXP--NAU   &	930 &	30 &	$5\cdot 10^{-19}$ &	$85.5^{\pm 0.5}_{}$ &	$0.028^{\pm 0.009}$ &	0 &	4.2\\
AUR--NAU   &	930 &	30 &	$1\cdot 10^{-18}$ &	$90.0^{\pm 0.5}_{}$ &	$0.081^{\pm 0.015}$ &	0 &	4.0\\
AUR--VIR   &	930 &	30 &	$2\cdot 10^{-18}$ &	$69.6^{\pm 0.7}_{}$ &	$0.106^{\pm 0.017}$ &	0 &	5.1\\
 \hline
AUR--NAU   &	938 &	30 &	$2\cdot 10^{-20}$ &	$4.6^{\pm 0.3}_{}$ &	$3.0^{\pm 0.09}$ &	6 &	214.1\\
AUR--NAU   &	938 &	30 &	$5\cdot 10^{-20}$ &	$41.3^{\pm 0.7}_{}$ &	$0.068^{\pm 0.014}$ &	0 &	8.6\\
AUR--NAU   &	938 &	30 &	$1\cdot 10^{-19}$ &	$68.1^{\pm 0.7}_{}$ &	$0.078^{\pm 0.014}$ &	0 &	5.2\\
AUR--VIR   &	938 &	30 &	$2\cdot 10^{-19}$ &	$63.0^{\pm 0.7}_{}$ &	$0.056^{\pm 0.012}$ &	0 &	5.6\\
AUR--NAU   &	938 &	30 &	$5\cdot 10^{-19}$ &	$86.9^{\pm 0.5}_{}$ &	$0.081^{\pm 0.015}$ &	0 &	4.1\\
AUR--EXP   &	938 &	30 &	$1\cdot 10^{-18}$ &	$86.3^{\pm 0.5}_{}$ &	$0.073^{\pm 0.014}$ &	0 &	4.1\\
AUR--VIR   &	938 &	30 &	$2\cdot 10^{-18}$ &	$70.2^{\pm 0.7}_{}$ &	$0.098^{\pm 0.016}$ &	0 &	5.1\\
\end{longtable}

\section{Final remarks}

We presented a methodological study for analyzing data collected by a network
of non-homogeneous detectors. The search was aimed at detecting transient GW
signals. We implemented a two-fold time coincidence search; however, this
method could be applied as well to any detectors combination, e.g. three-fold,
logical ``OR'' of two-fold, etc.. For each set of waveform parameters and
amplitude, 6 different couples of detectors were available: we chose to
perform our search on those couples which allow potentially the set up of the
most stringent upper limit (see sect. 5.4). 

The key point of the method is the optimization process of the analysis thresholds for 
a given source population by means of Monte Carlo MDC injections.

Although the proposed methodology is viable for any specific signal
model, including the sky distribution of the sources, in the preset
study we assumed DS signals incoming from the GC, limiting our
observation range to our galaxy.

Moreover, in order to estimate the detection efficiency of the network we applied a standard Monte Carlo procedure based on a large set of injected signals.
Although this software technique is computational intensive, it permits to
derive reliable values of the efficiency and unbiased physical interpretation
of the results, i.e. the GW amplitudes and rates of the population under
study. 

Finally we notice that in the procedure presented here, the
statistical test relies just on the event magnitude. However, we
stress that it is possible to extend the method either by including
other statistical tests in the definition of the local Event Trigger Generators or by
implementing a common maximum likelihood estimator ($\chi^{2}$ test).

\section{Acknowledgments}

We acknowledge the funding from EU FP6 programme - ILIAS. Lucio Baggio was
supported by the EGO Consortium. Virginia Re was supported by EGO-VESF fellowship, call 2005.

\appendix

\section{Optimization pipeline}

We present here a schematic overview of all the steps:
\begin{itemize}
\item 
A specific search algorithm for each detector is run in order to produce lists of triggers. For each trigger the time and an estimate of the SNR ratio are exchanged. The exchanged lists comprise one derived by the analysis of the plain data from the detector, and others obtained by adding to the data different MDC channels before running the event search algorithm.

\end{itemize}

\vspace{.7cm}
\paragraph{\textbf{Coincidence search}}
\begin{itemize}
 
\item
For each couple of detectors, the plain data sets are searched for
coincidences after adding 800 offsets in steps of 1s, with a time window of
40ms. This covers about 7 minutes before and after the unshifted time, with a
safety range of $\pm 20s$ around the zero-delay time\footnote{We recall that
  10s is the maximum value for the blind shift applied previously to data
  exchange, see section ~\ref{sec:data}.}.
\item
All the lists with MDC injections are searched for coincident events (obviously no time delay analysis is performed in these cases) and the detection efficiency is thus evaluated. 
\end{itemize}

\vspace{.7cm}
\paragraph{\textbf{Optimization procedure}}
\begin{itemize}
\item
The data are split into two equal sets, one used in the optimization phase, the other in the estimation phase, by alternating shift index.
\item
The data are divided in 30 minutes long time bins .
\item
We evolve a couple of staircase thresholds (jointly for the two detectors) by increasing the threshold level in one time bin and one detector at a time. At each step, in order to have a significant variation, the test threshold is increased by an amount which corresponds to a reduction in the background counts of the order of the standard deviation of the counts themselves\footnote{We found empirically that the square root of the counts divided by 6 is a good compromise. Yet, we impose that the background decrease of at least 4 counts.}.
\item
At the $n$th step of the algorithm, we compute the ratio
$N_{eff}^{(n)}/\sqrt{N_b^{(n)}}$, where $N_{eff}^{(n)}$ and $N_b^{(n)}$ are
the \textit{total} number of MDC coincidences and background coincidences from
the set reserved for the optimization whose associated amplitudes are above
the $n$th set of thresholds. Then the effect of increasing the threshold of
one level at one bin is evaluated by computing the new ratio
$N_{eff}^{(n+1)}/\sqrt{N_b^{(n+1)}}$. Every time the bins in both detectors are
tried one by one in order to find for which a threshold change would score the
higher benchmark. If this benchmark is better than the one obtained at the
previous step, the level for that bin is changed, and this is taken as the
starting point for the next loop. Instead, if all changes resulted in a
decrease of the benchmark, the loop is exited. The loop will otherwise
continue up to reaching the higher level of thresholds. 
\item
When a time bin is found contributing to the efficiency by less than $0.1\%$ the threshold in that bin is raised until no background event is surviving. Similarly to what was done, at the end of the previous loop we recompute the benchmark $N_{eff}/\sqrt{N_{b}}$ to see whether removing the bin would be an improvement. These actions mitigate the possibility that the algorithm converges to a false maximum with relatively high background.
\item
The found set of thresholds are applied to the alternative sets of triggers which were kept aside in order to re-estimate unbiased values for $N_{eff}$ and $N_{b}$. These two numbers, divided respectively by the number of injected events (4320) and of time-shifted configurations (400), give the estimates for the efficiency of detection and for the average background counts.
\end{itemize}

\vspace{.7cm}
\paragraph{\textbf{Efficiency and confidence}}
\begin{itemize}

\item
For each shifted configuration of the alternative set, the number of coincident events is used to compute the corresponding upper limit at 95\% confidence.
\item
This entire procedure is repeated for a different couple of detectors and/or a different set of MDC injections.
\item
After having decided the level of confidence for the eventual rejection of the Null hypothesis, we unveil the coincidence counts at zero-delay, and compute for them the confidence interval.

\end{itemize}


\section{Details about the injected signals}
\label{sec:MDCdetails}
In this study we assume that the waveforms, in the TT gauge
are of the form

\begin{eqnarray}
\left(\begin{array}{c}
h_{+}\nonumber\\
h_{\times}\end{array}\right)=&&\frac{h_{rss}}{\pi f_{gw}\tau}\sqrt{\frac{1+4\pi^{2}f_{gw}^{2}\tau^{2}}{\tau\left(1+e^{-1/\left(2f_{gw}\tau\right)}\right)}}e^{-t/\tau}\nonumber\\
&& \times \left(\begin{array}{cc}
\cos2\psi & -\sin2\psi\\
\sin2\psi & \cos2\psi\end{array}\right) \nonumber\\
&& \times \left(\begin{array}{cc}
\frac{1+\cos^{2}\iota}{2} &  \Theta(t-\frac{1}{4f_{gw}}) \cos\left(2\pi f_{gw}t\right) \\
\cos\iota & \Theta(t) \sin\left(2\pi f_{gw}t\right)\end{array}\right)
\label{eq:formula}
\end{eqnarray}

where the angle $\psi$ is an arbitrary polarization and the angle
$\iota$ an arbitrary inclination of the angular momentum of the system which originates the burst with
respect to the line of sight.

Notice the different $\Theta$ for the $h_{+}$ and $h_{\times}$ terms;
the reason for multiplying $\cos\left(2\pi f_{gw}t\right)$ by
$\Theta\left(1-\frac{1}{4f_{gw}}\right)$ is to avoid a discontinuity at the
beginning of the waveform, which would result into an infinite energy, even
though $h_{rss}$ would remain finite.

The polarization $\psi$ is uniformly distributed in $[0,\,2\pi)$,
while $\cos i$ is uniformly distributed in $[-1,\,1)$; these choices
correspond to assume a random orientation in space of the axis of
symmetry of the emitting system.

\subsection{Signal normalization}
\label{norm}
The signal normalization is done requiring equation (B.2) for $\psi=0,\,\iota=0$:
\begin{equation}
h_{rss}^{2}\equiv\int_{0}^{\infty}\left(\left|h_{+}(t)\right|^{2}+\left|h_{\times}(t)\right|^{2}\right)dt\,.
\end{equation}

It can also be useful to 
relate $h_{rss}$ and the energy emitted $E$ assuming a source located at a distance
$r$. To this end, we recall the standard definition of the energy
flux
\begin{equation}
\frac{dE}{dAdf}=\frac{\pi c^{3}}{2G_{N}}f^{2}\left(\left|\tilde{h}_{+}(f)\right | ^{2}+\left|\tilde{h}_{\times}(f)\right| ^{2}\right)
\end{equation}

 where $dA=r^{2}d\Omega$. It is straightforward to compute

\begin{eqnarray}
&& \left(\left|\tilde{h}_{+}(f)\right|^{2}+ \left|\tilde{h}_{\times}(f)\right|^{2}\right)= h_{rss}^{2}\frac{\tau  \left(4 f_{gw}^2 \pi^2 \tau^2+1\right)}{\left(1+e^{\frac{1}{2 f_{gw} \tau }}\right)} \nonumber\\ 
&&\times\frac{\left(\cos^4(i)+\left(2+4 e^{\frac{1}{2 f_{gw} \tau }}\right) \cos^2(i)+1\right)}{ \left(16 f^4 \pi^4 \tau^4+\left(4 f_{gw}^2 \pi^2 \tau^2+1\right)^2+f^2 \left(8 \pi^2 \tau^2-32 f_{gw}^2 \pi^4 \tau^4\right)\right)}
\label{eq:ap}
\end{eqnarray}

and then, after performing the integral over frequencies and the angles, one obtains:
\begin{equation}
E=\frac{\pi c^{3} r^{2} h_{rss}^{2} \left(4 f_{gw}^{2} \pi^{2} \tau^{2}+1\right)}{2 G_{N}}  \frac{\left(7+5 e^{\frac{1}{2 f_{gw} \tau }}\right) }{30\left(1 + e^{\frac{1}{2 f_{gw} \tau}}\right) \pi \tau^{2}}
\end{equation}

i.e.:
\begin{eqnarray}
\frac{E}{M_{\odot}c^{2}}\simeq&&10^{-6}\frac{\left(1+(5/7) e^{\frac{1}{2
      f_{gw} \tau }}\right)}{\left(1+e^{\frac{1}{2 f_{gw} \tau }}\right)}\left(1+\frac{1}{4\pi^{2}\tau^{2}f_{gw}^{2}}\right) \nonumber\\
 &&\times \left[\frac{h_{rss}}{10^{-21}/\sqrt{Hz}}\right]^{2}\left[\frac{r}{10\,kpc}\right]^{2}\left[\frac{f_{gw}}{1kHz}\right]^{2}
\end{eqnarray}

which means that observing an $h_{rss}\simeq10^{-21}/\sqrt{Hz}$ with signals
at about $1$kHz corresponds to a source emitting 
a fraction $\sim 10^{-6}$ 
of a solar mass in gravitational waves at a distance of $10$kpc.

\begin{center}
 \section*{References}
\end{center}



\begin{thebibliography}{99}
\expandafter\ifx\csname natexlab\endcsname\relax\def\natexlab#1{#1}\fi
\expandafter\ifx\csname bibnamefont\endcsname\relax
  \def\bibnamefont#1{#1}\fi
\expandafter\ifx\csname bibfnamefont\endcsname\relax
  \def\bibfnamefont#1{#1}\fi
\expandafter\ifx\csname citenamefont\endcsname\relax
  \def\citenamefont#1{#1}\fi
\expandafter\ifx\csname url\endcsname\relax
  \def\url#1{\texttt{#1}}\fi
\expandafter\ifx\csname urlprefix\endcsname\relax\def\urlprefix{URL }\fi
\providecommand{\bibinfo}[2]{#2}
\providecommand{\eprint}[2][]{\url{#2}}


\bibitem{3-modes} 
 Baggio L \textit{et al}
\newblock 2005 {\em Phys.\ Rev.\ Lett.} \textbf{94} 241101
\bibitem {AURIGA} Vinante A (for~the AURIGA~Collaboration)
\newblock 2006 {\em Class.\ Quantum\ Grav.} \textbf{23} S103
\bibitem{ROGa}
\bibinfo{author}{\bibnamefont{Astone}~\bibfnamefont{P}} 
  \bibinfo{author}{\bibnamefont{(for~the ROG~Collaboration)}}
  2004 {\em Class.\ Quantum\ Grav.}
  \textbf{\bibinfo{volume}{21}}
  (http://stacks.iop.org/0264-9381/21/S1585)

\bibitem{ROGb}
\bibinfo{author}{\bibnamefont{Astone}~\bibfnamefont{P}},
  \textit{et~al} 2006 {\em Class.\ Quantum\ Grav.}
  \textbf{\bibinfo{volume}{23}} \bibinfo{pages}{S57}
  (http://stacks.iop.org/0264-9381/23/S57)
\bibitem{Virgo} Acernese F (for the Virgo collaboration) 2006 {\em Class.\
  Quantum\ Grav.} {\bf 23} S635 and referencies therein

\bibitem{IGEC-PRL} Allen B Z (for the IGEC Collaboration) 2000 {\it Phys. Rev. Lett.} {\bf 85} 5046
\bibitem{IGEC-PRD} Astone P \textit{et al} (IGEC Collaboration) 2003 \textit{Phys. Rev. D} \textbf{68} 022001
\bibitem{ROG} Astone P \textit{et al} 2002 {\em Class.\
  Quantum\ Grav.} {\bf 19} 5449

\bibitem{LIGO2004} Abbott B \textit{et al} (LIGO Scientific Collaboration) 2004 \textit{Phys. Rev. D} \textbf{69} 102001
\bibitem{LIGO2005} Abbott B \textit{et al} (LIGO Scientific Collaboration) 2005 \textit{Phys. Rev. D} \textbf{72} 062001
\bibitem{LIGO2006} Abbott B \textit{et al} (LIGO Scientific Collaboration) 2006 \textit{Class. Quantum Grav.} {\bf 23} S653

\bibitem{LIGO-TAMA-S2} Abbott B \textit{et al} (LIGO Scientific Collaboration)
  2005 
\textit{Phys. Rev. D} {\bf72} 122004


\bibitem{AURLIGOgwdaw} Cadonati L \textit{et al} 2005 \textit{Class. Quantum Grav.} \textbf{22} S1337
\bibitem{AURLIGOamaldi6} Poggi S, Salemi F (for the AURIGA collaboration)
  and Cadonati L (for the LIGO Scientific collaboration) 2006
  \textit{J. Phys.: Conf. Ser.} \textbf{32} 198

\bibitem{BH}  
  Kokkotas K D and Schmidt B G 1999 http://www.livingreviews.org/lrr-1999-2
\bibitem{f-modes} Ferrari V \textit{et al} 2003 \textit{Mon. Not. Roy. Astron. Soc.} {\bf 342} 629
\bibitem{PF} Guidi G M, Cuoco E and Vicer\'e A 2004 \textit{Class. Quantum Grav.} {\bf 21} S815
\bibitem{Waveburst} S.~Klimenko \textit{et al} LIGO-T050222-00-Z
\bibitem{ROG1} Astone P {\it et al.} 1997 {\it Il Nuovo Cimento} {\bf 20}
\bibitem{ROG2} D'Antonio S 2002 \textit{Class. Quantum Grav.} {\bf 19} 1499
\bibitem{siesta:98} Caron B, Flaminio R, Marion F, Mours B, Verkindt D,
  Cavalier F and Vicer\'e A 1999 \textit{ Astrop. Phys.} {\bf 10} 369
\bibitem{Drago-efficiency} Drago M, private communication
\bibitem{ROG3} Astone P, D'Antonio S and Pai A 2006 \textit{J. Phys.: Conf. Ser.} {\bf 32} 192 

\bibitem{feldmancousinsPRD} Feldman G J and Cousins R D 1998 
\textit{Phys. Rev. D} {\bf 57} 3873
\bibitem{PHYSTAT2003} Baggio L and Prodi G A 2003 \textit{Statistical problems
  in particle physics, astrophysics and cosmology} ed R Mount, L Lyonsand and
  R Reitmeyer (Stanford) 238
  (http://www.slac.stanford.edu/econf/C030908/papers/WELT003.pdf, arXiv:astro-ph/0312353)


\end{thebibliography}
\end{document}